# Carbon nanotube-copper fibers produced by electrospinning: electrical conductivity measurements

## Farhad Daneshvar[1], Hung-Jue Sue

*Department of Materials Science and Engineering, Texas A&M University, College Station, TX 77843, USA.*

## 1. Background and motivation

Subsea power transmission cables are taking on greater significance in the smart grids that are set to overhaul the existing power infrastructure. According to Navigant research the market for high voltage cables alone projected to surpass 5.3 billion USD in annual revenue by 2023, up from 1.4 billion in 2014 [1].

Submarine cables are mainly made of copper or aluminum alloys. Pure copper has a very low electrical resistivity however it suffers from poor mechanical properties and high density which has challenged its application in subsea cables. Aluminum and its alloys on the other hand have higher specific conductivity (=conductivity/density) and generally better mechanical properties but for achieving the same power transmission performance the diameter of the cable should be increased. As a result of the added bulk, the bending radius increases which leads to many instalment challenges.

Recent advances in nanotechnology have provided new materials which have the potential to surpass copper and aluminum alloys in electrical conductivity, weight and ampacity [2-6]. Among these carbon nanotubes (CNTs) stand out due to their remarkable thermal and electrical conductivity and ampacity ($10^3$ times of copper), low density, abundance of precursor materials and supreme mechanical properties. However, making these materials into a continuous fiber or macrostructures has remained the main obstacle.

A promising approach to tackle this issue is employing CNTs as nanofillers in copper matrixes. Subramaniam et al. [5] combined several simple fabrication to produce high density CNT (45 vol%)-Cu composite films with specific conductivity 26% greater than copper. These results show that by improving the interfacial interactions between the CNTs and the copper matrix strong and highly conductive nanocomposites can be made. In this research continuous CNT-Cu nanocomposite is produced using electrospinning method. For improving the interfacial interaction the surface of the CNTs were first coated with a thin layer of copper using electroless deposition methods (the process is explained in details in [7]).

## 2. Experimental procedure

0.2 g Cu coated CNTs (Cu/CNTs (Figure 1(a)) were dispersed in 10 ml aqueous solution of PVA (8 wt%) at 80 °C. Subsequently 0.2 g copper acetate was added to the solution above and the mixture was stirred as it cooled down to the room temperature. This mixture was used in electrospinning to produce CNT-Cu-PVA fibers. The shear stress caused by the needle aligns the CNTs in the direction of the fiber and increases the density. The fibers were heated at 330 °C in air to remove the PVA and then in hydrogen at 400 °C for 6 h to reduce the fibers to Cu-CNT. The morphology and composition of the fibers were studied using scanning electron microscopy (SEM) equipped with Energy-dispersive X-ray spectroscopy (EDS) Fibers were attached to standard microchips using focused ion beam (FIB) to measure the electrical conductivity (FEI FIB200 system at 30 keV and 13 pA).

---

[1] Corresponding author: Farhad Daneshvar Email: f.daneshvar@tamu.edu

# 3. Results and discussion

As a result of electrospinning the Cu and CNT-rich solution will accumulate at the center of the fiber and the polymer-rich solution will cover it forming a core/shell structure. By removal of the PVA in air at 330 °C CNT-CuO fiber with a coarse surface (Figure 1 (b)) is achieved. It has been reported before that CNt introduction to metals can increase the roughness of the composites [8]. It should be noted that since electrons migrate through the surface of conductive materials the resurface roughness may deteriorate the electrical conductivity of the fibers. Hence it is required to do a heat treatment in hydrogen environment at 400°C to not only reduce the CuO to Cu but also decrease the surface roughness (Figure 1(c)). As it can be observed in Figure 1(d), EDS result show that the fiber is made of pure copper. Since the CNTs were covered with a thin layer of Cu prior to electrospinning detecting individual CNTs inside the Cu matrix is very challenging.

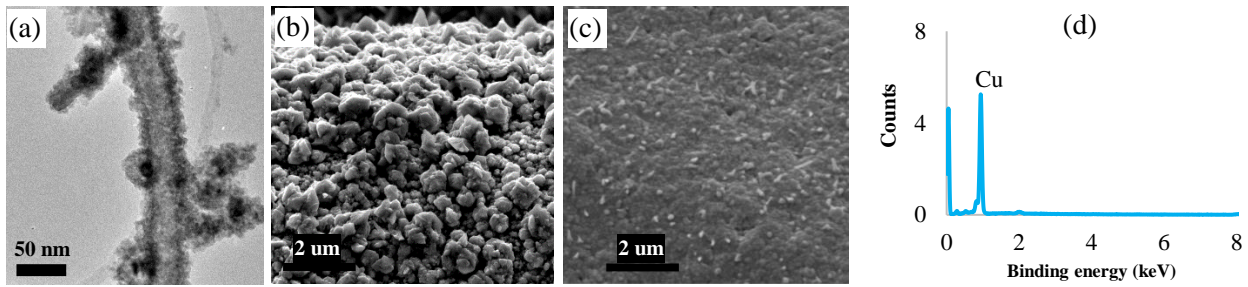

*Figure 1. (a) TEM of a Cu coated CNT. (b) SEM image of CNT-Cu fiber before heat treatment and (c) after heat treatment. (d) EDS spectrum of CNT Cu fiber. Cu peak is clearly observed. No impurity was observed in the EDS spectrum. Small peaks of C and O were also detected before Cu peak that can be due to SEM chamber contamination.*

For measuring the conductivity of CNT-Cu nanofibers a gold pad was prepared using e-beam evaporator and lift off process. Next the Cu-CNT fiber drop-casted on a polydimethylsiloxane (PDMS) tape is located on the gold pad and the tape is removed. This process is schematically summarized in Figure 2(a). Due to nano size of these wires handling and testing was done in the SEM chamber. Also to establish the lowest possible contact resistance [2], a few nanometers of the Cu-CNT was etched using gallium focused ion beam (Ga-FIB) milling and then about 200 nm thick, 1 μm wide and 5 μm long platinum (Pt) pads were deposited using a Ga-FIB (Figure 2(b)). It should be noted that during the conductivity measurement the beam was shut off since it can affect the result. As presented in Figure 2(c) the specific conductivity of the fiber is comparable to copper showing the feasibility of production continuous CNT-Cu fibers with high electrical properties.

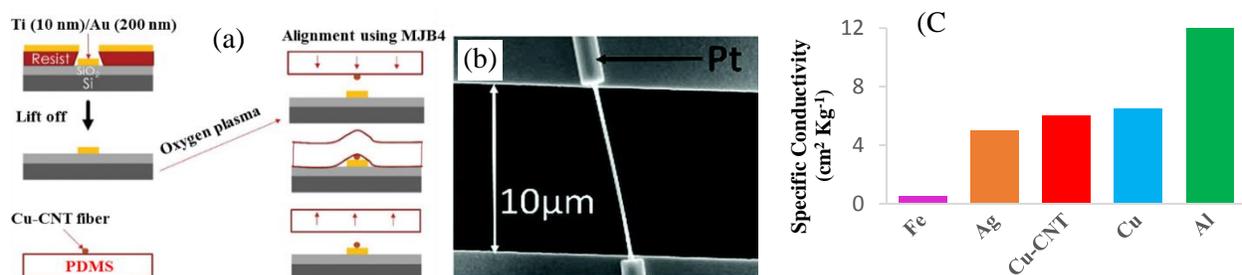

*Figure 2. (a) Schematic of the Cu-CNT fiber transfer process from PDMS to the Ti/Au electrode on a Si/SiO2 wafer; SEM image; of the device after Pt deposition using a Ga focused ion beam  Comparison of conductivity per unit weight (specific conductivity) of Cu-CNT with different metals.*

**Conclusion**

For the first time continuous Cu-CNT fibers were made using electrospinning method. For enhancing the interaction between the CNT and the Cu matrix, the CNT surface was coated with copper using electroless deposition method. In the last step a heat treatment step was performed on the fibers at 330 °C in air and 400 °C in hydrogen to modify the morphology and composition of the fibers. By the aid of FIB-SEM electrical conductivity of the individual fibers were measured. Results showed that the specific conductivity of the composite fiber can reach to 5.9 cm$^2$ Kg$^{-1}$ which is comparable to copper. By considering the effect of CNTs in improving the mechanical properties of metals [3, 9-11], these results show that electrospinning is a promising technique for fabrication of a new generation of materials for power transmission applications.